\documentclass[final,english]{bullsrsl}[2022/06/15]

\RequirePackage{hyperref}

\usepackage[latin1]{inputenc}
\usepackage[T1]{fontenc}
\usepackage{natbib} 
\usepackage{graphicx}
\usepackage{subcaption}

\begin{document}
\title{Hot Stellar Populations of Berkeley 39 using \textit{Swift}/UVOT}

\author[affil={}, corresponding]{Komal}{Chand}
\author[affil={}]{Khushboo Kunwar}{Rao}
\author[affil={}]{Kaushar}{Vaidya}
\author[affil={}]{Anju}{Panthi}

\affiliation[]{Department of Physics, Birla Institute of Technology and Science-Pilani, 333031 Rajasthan, India}

\correspondance{p20210463@pilani.bits-pilani.ac.in}
\date{17th May 2023}
\maketitle

\begin{abstract}
Open clusters are excellent tools to probe the history of the Galactic disk and properties of star formation. In this work, we present a study of an old age open cluster Berkley 39 using the observations from UVOT instrument of the Neil Gehrels Swift observatory. Making use of a machine learning algorithm, ML-MOC, we have identified a total of 861 stars as cluster members out of which 17 are blue straggler stars. In this work, we present a characterisation of 2 blue straggler stars. To estimate the fundamental parameters of blue straggler stars and their companions (if any), we constructed spectral energy distributions using UV data from \textit{swift}/UVOT and GALEX, optical data from \textit{Gaia} DR3, and infrared (IR) data from 2MASS, \textit{Spitzer}/IRAC, and WISE. We find excess flux in UV in one blue straggler star, implying the possibility of a hot companion.
\end{abstract}


\section{Introduction}
Open clusters (OCs) showcase a wide variety of stellar populations providing a glimpse into the formation and evolution of stars. They contain hundreds to thousands of gravitationally bound stars formed in the same star formation event from the same parent giant molecular cloud (GMC, \citealt{Harris2009}). Therefore, member stars of OCs are a group of stars with similar ages, distances, and metallicities. Dynamical interactions in OCs can lead to the formation of exotic stellar populations such as blue straggler stars (BSS, \citealt{sandage1953}), yellow straggler stars \citep{Strom1971, Leiner2016}, red straggler stars \citep{Geller2017}, and cataclysmic variables \citep{Ritter2010}, that challenge the standard model of single star evolution.
BSS are rejuvenated main-sequence stars that are brighter and bluer than the main-sequence turnoff (MSTO) in the color-magnitude diagrams (CMD). Their formation mechanisms and evolution processes are still debatable and are an active field of research \citep{Boffin2015}. The widely accepted formation mechanisms of BSS are: (a) Mass transfer (MT) process: this occurs when a BSS is in a binary system with a companion star, and this companion star evolves and expands, (b) Merger process: this occurs when two stars are in close proximity and they merge due to loss of angular momentum to form a more massive star \citep{Andronov2006}, and (c) Collision process: this occurs because of direct stellar collisions leading to mergers \citep{Hills1976, Leonard1996}. Direct stellar collisions are not feasible for low-density environments such as OCs \citep{Chatterjee2013}. Based on the evolutionary stage of the primary star, MT mechanism is further divided into three categories: Case A -- primary star is in the main sequence phase of its evolution \citep{Webbink1976}, Case B -- primary star is in the red-giant branch (RGB) phase \citep{McCrea1964}, and Case C -- primary star is in the asymptotic giant branch (AGB) phase \citep{Chen2008}.

With an objective of characterizing the BSS population, determining their properties, and detecting their hot companions, we studied Berkeley 39 using archival data from \textit{Swift}/UVOT and other wavelengths. Berkeley 39 is a 6 Gyr old \citep{Kassis1997} massive OC ($\sim$2$\times 10^{4} \, M_{\odot}$, \citealt{Lata2002}) having a mean metallicity [Fe/H] = $-$0.20 \citep{Bragaglia2012}. It is located at a distance of $\sim$3.8 kpc. The BSS population of this cluster was extensively studied by \cite{Vaidya2020} and \cite{Rao2021}. \cite{Vaidya2020} identified 23 BSS on the basis of the magnitude ($G \leq  MSTO_{Mag} + 0.5$) and color ($BP -RP \leq MSTO_{Color} - 0.05$) of the MSTO \cite{Rao2021} identified 16 BSS candidates based on the fitted zero-age main sequence, parsec isochrone, and equal mass binary isochrone (G = 0.75 mag brighter than the fitted parsec isochrone). \cite{Vaidya2020} and \cite{Rao2021} found this cluster to be dynamically young or intermediate age on the basis of the segregation of its BSS with respect to other less massive reference populations.  With a reasonably large number of BSS, this cluster is an ideal cluster to study in ultraviolet (UV) wavelengths. Although the BSS of this cluster have been studied to understand the dynamical status of the cluster by \citet{Vaidya2020} and \citet{Rao2021}, the formation mechanisms of the BSS have not been studied earlier. We found \textit{Swift}/UVOT archival data for this cluster in three near-UV filters, \textit{UVW1}, \textit{UVW2}, and \textit{UVM1}. Combining UVOT data with photometric data in other filters provides valuable insights into the fundamental physical properties of astronomical objects by characterizing their emissions across a wide range of wavelengths. Since BSS are hot, they radiate a significant amount of their flux at UV wavelengths. Additionally, BSS may exhibit UV excess owing to the presence of an unresolved hot companion. Therefore, spectral energy distributions (SEDs) covering a long wavelength range from far UV to far IR bands, have the potential to explain the formation and evolution of BSS along with their hot companions \citep{Gosnell2015, Subramaniam2016, subramaniam2018ultraviolet, Sindhu2019, 2022MNRAS.511.2274V, Rao2022, Panthi2022}.

\section{Observation and Data Analysis}

\subsection{Optical Data}
We used optical data from \textit{Gaia} DR3 {\citep{2021A&A...649A...1G}} to identify the cluster members using a machine-learning based membership determination algorithm for open clusters (ML-MOC, \citealt{Agarwal2021}). ML-MOC uses k-Nearest Neighbour (kNN, \citealt{Cover1967}) and Gaussian mixture model (GMM, \citealt{Peel2000finite}). To identify cluster members, this algorithm uses parallax and proper motion from \textit{Gaia} DR3 data. We identified a total of 861 sources as cluster members, out of which 17 are classified as BSS. The details of the two BSS that we present in this work are provided in Table \ref{Table1}.

\begin{table}
\label{Table1}
\centering
\caption{\footnotesize Basic details of two BSS presented in this work. }
\resizebox{\textwidth}{!}
{
\begin{tabular}{cccccccc}
\hline
\multicolumn{1}{c}{Name} & \multicolumn{1}{c}{RA} & \multicolumn{1}{c}{DEC} & \multicolumn{1}{c}{GAIA DR3 source\textunderscore id}  & \multicolumn{3}{c}{UV flux (ergs s$^{-1}$ cm$^{-2}$ \AA$^{-1}$)} \\
 &  &  &  & UVW2  & UVM2 & UVW1\\
\hline
BSS3 & 116.74244 & $-$4.69012 & 3056678441602599296 & 9.83405e-16  & 7.79188e-16 & 6.964300e-16\\
BSS9 & 116.61005 & $-$4.66228 & 3056682289893334912 & 2.40237e-17  & 2.09187e-17 & 8.136015e-17\\
\hline
\end{tabular}
}
\label{Table1}
\end{table}

\subsection{UV Data}
UV data are taken from \textit{Swift}/UVOT and Galaxy Evolution Explorer (GALEX). The UVOT is a 30 cm modified Ritchey-Chretien UV/optical telescope co-aligned with the X-ray telescope and having a wide-field view of $17^{\prime} \times 17^{\prime}$. The UVOT instrument covers a wavelength range of 170 -- 650 nm. It utilises clear white filters, U (300 -- 400 nm), B (380 -- 500 nm), V (500 -- 600 nm), UVW1 (220 -- 400 nm), UVM2 (200 -- 280 nm), and UVW2 (180 -- 260 nm), two grisms, a magnifier, and a blocked filter \citep{2005SSRv..120...95R}. \textit{Swift}/UVOT observed  Berkeley 39 in three UV filters, UVW1, UVM2, and UVW2 in 2011. A Catalog of this cluster for near ultra-violet (NUV) point source is provided by \cite{Siegel2019}. GALEX is a space-based telescope \citep{Martin2005} that operates in two UV bands: far-UV (1350 -- 1780 \AA) and NUV (1770 -- 2730 \AA).

\subsection{Infrared data}
Infrared data (IR) are taken from two Micron All-Sky Survey (2MASS), \textit{Spitzer}/IRAC, and Wide-field Infrared Survey Explorer (WISE).
2MASS \citep{Cohen2003} provides a near-IR survey of the entire sky in the J-band (1.235 $\mu$m), H-band (1.662 $\mu$m), and Ks-band (2.159 $\mu$m). \textit{Spitzer}/IRAC catalog \citep{fazio2004infrared} contains information  about sources detected in all four filters I1 (3.6 $\mu$m), I2 (4.5 $\mu$m), I3 (5.8 $\mu$m), and I4 (8.0 $\mu$m). Berkeley 39 was observed in all four IRAC channels in 2006 with an exposure time of 26.8 s. WISE is a mid-IR full sky survey which contains four different bands, W1, W2, W3, and W4, with wavelengths centered at 3.35 $\mu$m, 4.60 $\mu$m, 11.56 $\mu$m, and 22.09 $\mu$m, respectively \citep{Wright2010}. For sources with no IRAC magnitudes in the I1 and I2 channels, we use the WISE W1 and W2 band fluxes, whereas WISE W3 and W4 band fluxes are used for all the sources when available. When only upper limits are available in WISE W3 and W4 bands, they are shown on the plots but not included in the fits.

\section{Analysis of BSS}
\begin{figure*}
        \centering
    \begin{subfigure}[t]{0.49\textwidth}
       
       \includegraphics [width=1.0\textwidth]
       {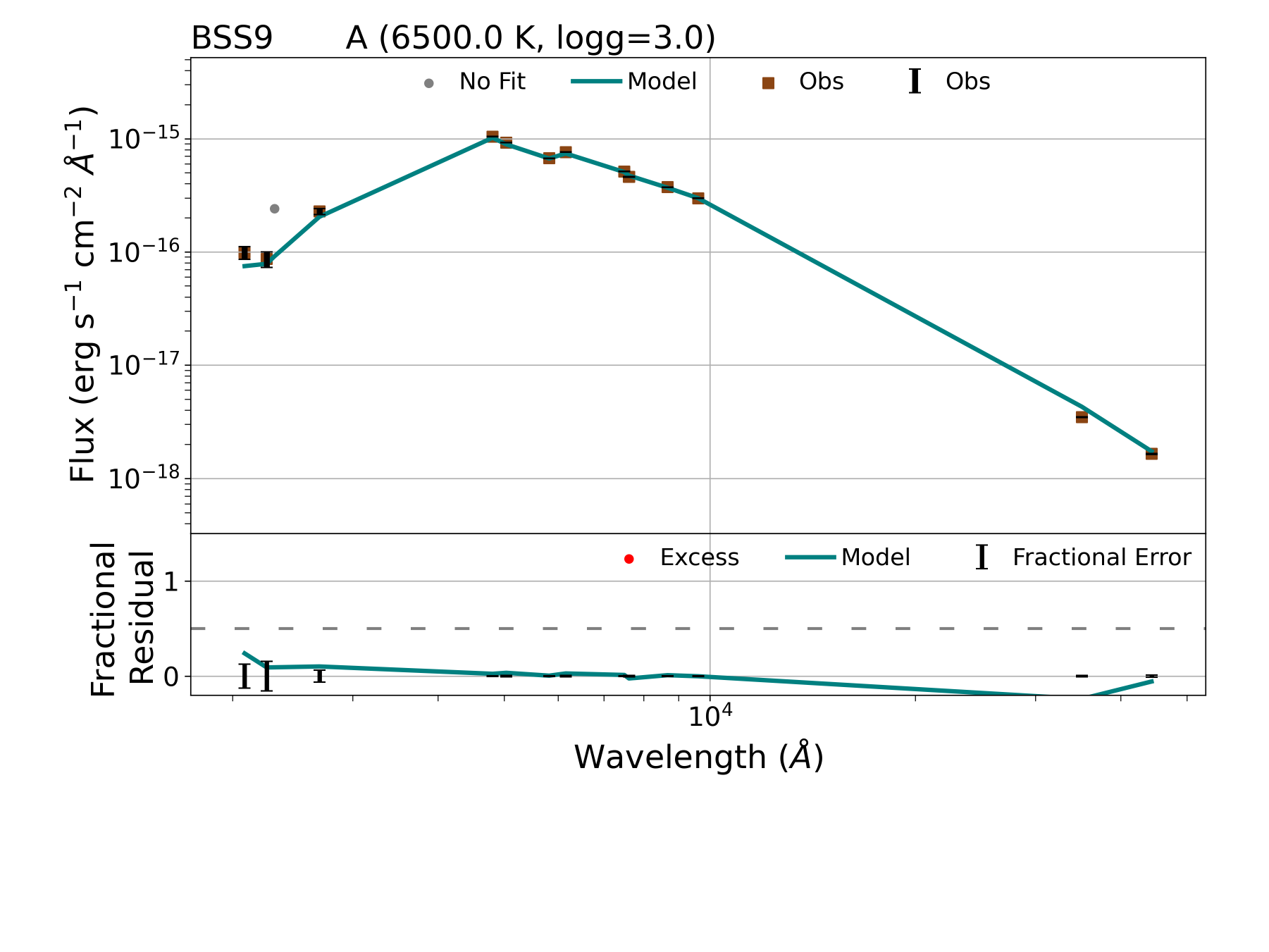}
       \vspace{-1.45cm}
       \caption*{(a)}
    \end{subfigure}
    \hfill
    \begin{subfigure}[t]{0.49\textwidth}
    
       \includegraphics [width=1.0\textwidth]
       {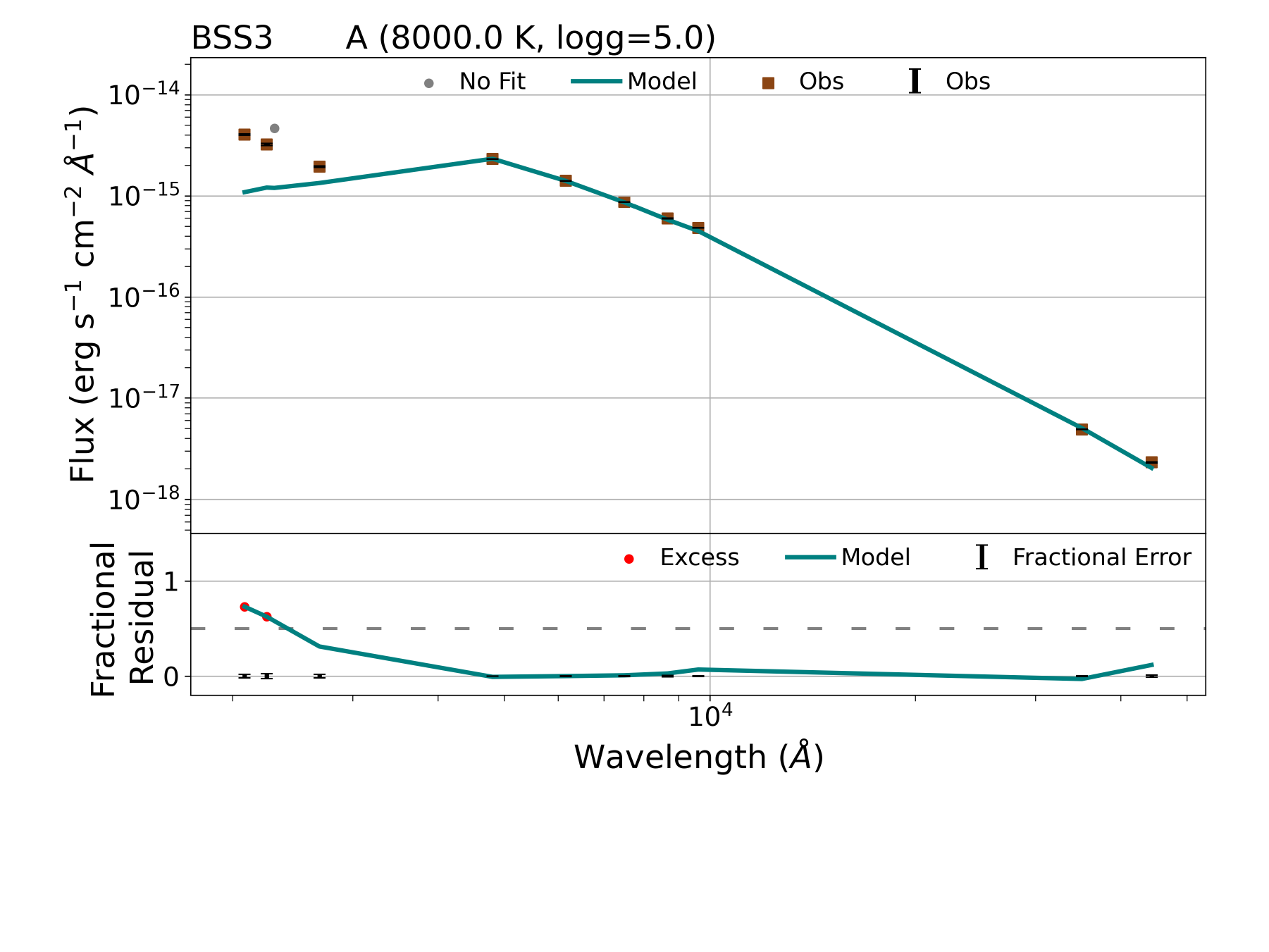}
       \vspace{-1.45cm}
       \caption*{(b)}
    \end{subfigure}
    \hfill
   \caption{\footnotesize (a) The single component SED fit of BSS9. (b) The single component SED fit of BSS3. The top panels in each figure show the extinction corrected observed fluxes with brown data points, error bars representing the errors in observed fluxes in black, and the blue curve representing the Kurucz stellar model fit. The bottom panels depict the residual between extinction-corrected observed fluxes and the model across the filters from UV to IR wavelengths. BSS3 shows a fractional residual greater than 0.5 in all three UV filters.}
    \label{Figure1}
\end{figure*}

 We identified a total of 17 BSS, out of which 16 BSS had no nearby sources within 3$^{\prime \prime}$ radial distance and were analysed using SEDs. In this work, we focus on two BSS, BSS3 and BSS9. To construct the SEDs, we used the Virtual Observatory SED Analyzer (VOSA, \citealt{Bayo2008}). We provide VOSA with an input file which contains BSS coordinates, UV fluxes from \textit{Swift}/UVOT and GALEX, mid-IR fluxes from \textit{Spitzer}/IRAC, mean distance = 3.8 kpc, and extinction in the V-band ($A_{\mathrm{V}}$) as 0.51 mag. We estimated ($A_{\mathrm{V}}$) and the mean distance of the cluster using the bright cluster members (G $\leq$ 16 mag) referring to the \cite{Bailer2021} catalog for conversion from parallaxes to distances of individual members. VOSA then assembles fluxes in optical data from \textit{Gaia} DR3 and Pan-STARRS \citep{Chambers2016}, near-IR from 2MASS \citep{Cohen2003}, and mid-IR data from WISE \citep{Wright2010}. In order to normalise the SEDs, VOSA uses the distance and $A_{\mathrm{V}}$. We first excluded the UV data points from the SEDs to ensure that optical and IR data points were fitting satisfactorily with the model flux. We examined if there were any noticeable excess in the UV and/or IR data points and noted down the residuals in all the UV filters. After that, we used Kurucz stellar model \citep{Castelli1997} to fit the extinction-corrected observed fluxes to construct the SEDs. We used temperature T$_{\mathrm{eff}}$ and log\,\textit{g} as free parameters, with log\,\textit{g} ranging from 3 to 5 and T$_{\mathrm{eff}}$ from 3500 to 50000 K, whereas fixed the value of metallicity $([Fe/H])$ to be zero since it is closest to the cluster metallicity, $-$0.20 \citep{Bragaglia2012}. VOSA uses the $\chi^2$ minimisation technique to choose the best-fit SEDs. The reduced $\chi^2$ ($\chi_r^2$) value is obtained using following expression:\\
\begin{equation}
      \chi_r^2=\frac{1}{N-N_f}\sum_{i=1}^{N}\frac{(F_{o,i}-M_dF_{m,i})^2}{\Sigma^{2}_{o,i}}\\
\end{equation}

where $N$ indicates the total number of photometric data points, $N_f$ is the total number of fitted model parameters, $F_{o,i}$ is the observed flux, $F_{m,i}$ is the model flux of the star and $M_d=(\frac{r}{d})^2$ is the scaling factor, which is dependent on the radius and the distance of objects.\\
BSS9 did not show UV excess, whereas BSS3 showed UV excess with a fractional residual greater than $0.5$ in all three UV filters. Figure \ref{Figure1}(a) shows \textit{single-component} SED \textit{fit} of BSS9 whereas Figure \ref{Figure1}(b) shows \textit{single-component} SED \textit{fit} of BSS3. We tried to fit the UV excess in BSS3 by using the \textit{Binary$\_$SED$\_$fit} pipeline by \cite{Jadhav2021}. However, we were unsuccessful in finding a suitable model to fit the excess using Koester \citep{Koester2010} and Kurucz \citep{Castelli1997} models. Due to the unavailability of data in the XMM and \textit{Chandra} catalogs for this cluster, we cannot exclude the possibility of the presence of chromospheric activity as a cause of the UV excess in this BSS. Therefore, in order to gain more understanding of the nature of BSS3, we need further observations and revisit the SED fitting. Spectroscopic observations from the 3.6 m Devesthal Optical Telescope (DOT) can enable us to know the fundamental parameters of this BSS, which will help to constrain the SED and investigate the physical reason behind the UV excess. The analysis of the remaining BSS will be presented in Chand et al. (2023, in preparation).\\

\section{Summary}
We used the \textit{Swift}/UVOT data along with other multi-wavelength data to characterize two BSS of an old OC Berkeley 39. BSS3 shows an excess in the UV wavelengths, whereas BSS9 shows no UV excess. We are unsuccessful in fitting a suitable model to the SED of BSS3. As the UV excess could as well be due to reasons other than the presence of a hot companion, we searched the XMM, and \textit{Chandra} database for this BSS position. However, the X-ray data are not available for this region. Hence, we are unable to comment if this BSS shows UV excess likely due to unresolved hot companion or other reasons. We conclude that we need to revisit this BSS with better photometric and spectroscopic information to learn about its properties and detect the presence of any hot companions.

\begin{acknowledgments}
We thank the anonymous referee for their valuable comments. This work has made use of the third data release from the European Space Agency (ESA) mission {\it Gaia} (\url{https://www.cosmos.esa.int/gaia}), {\it Gaia} DR3 \citep{2021A&A...649A...1G}, processed by the {\it Gaia} Data Processing and Analysis Consortium (DPAC, \url{https://www.cosmos.esa.int/web/gaia/dpac/consortium}). This research has made use of the VizieR catalog access tool, CDS, Strasbourg, France. This research also made use of the Astrophysics Data System (ADS) governed by NASA (\url{https://ui.adsabs.harvard.edu}).

\end{acknowledgments}

\begin{furtherinformation}

\begin{authorcontributions}
Komal Chand -- Data curation, Formal analysis, Investigation, Methodology, Visualisation, Writting-original draft.\\
Khushboo Kunwar Rao -- Conceptualization, Data curation, Formal analysis, Validation, Writing-review and editing.\\
Kaushar Vaidya -- Investigation, Methodology, Resources, Supervision, Validation, Writting-review and editing.\\
Anju Panthi -- Data curation, Validation, Writing-review and editing.

\end{authorcontributions}
\begin{conflictsofinterest}
The authors declare no conflict of interest.
\end{conflictsofinterest}

\end{furtherinformation}
\bibliographystyle{bullsrsl-en}
\bibliography{extra}
\end{document}